\documentclass[english,useAMS,usenatbib]{mn2e} 
\usepackage{lmodern}

\usepackage[T1]{fontenc}
\usepackage[latin9]{inputenc}
\usepackage{color}
\usepackage[usenames,dvipsnames]{xcolor}
\usepackage{amsmath}
\usepackage{amssymb}
\usepackage{graphicx}
\usepackage{esint}
\usepackage{placeins}
\usepackage[authoryear]{natbib}

\usepackage{hyperref}
\definecolor{darkgreen}{rgb}{0.1,0.6,0.7}
\hypersetup{colorlinks = true,urlcolor=blue,citecolor=Blue}

\makeatletter
\usepackage{lmodern}\usepackage{epstopdf}

\newcommand{\half}{\frac{1}{2}}

\newcommand{\trace}[1]{\ensuremath{ \mathrm{Tr}\left( {#1} \right)   }}

\newcommand{\boX}{\ensuremath{\boldsymbol{X}}}
\newcommand{\bomu}{\ensuremath{\boldsymbol{\mu}}}
\newcommand{\boC}{{\sf{C}}}
\newcommand{\calW}{\ensuremath{\mathcal{W}}}

\newcommand{\boSig}{\ensuremath{{\sf{\Sigma}}}}
\newcommand{\boS}{\ensuremath{{\sf{S}}}}

\newcommand{\mathd}{\mathrm{d}}

\newcommand{\both}{\ensuremath{\boldsymbol{\theta}}}
\newcommand{\obsboX}{\ensuremath{\boX_{\mathrm{o}}}}

\newcommand{\boXim}{\ensuremath{\boX}}
\newcommand{\boXimbar}{\ensuremath{\bar{\boX}}}

\title[Parameter inference with estimated covariance matrices]{Parameter inference with estimated covariance matrices}
\author[E. Sellentin, A. F. Heavens]{Elena Sellentin$^{1}$, Alan F. Heavens$^{2}$\\
$^{1}$Institut f\"ur Theoretische Physik, Ruprecht-Karls-Universit\"at Heidelberg, Philosophenweg 16, 69120 Heidelberg, Germany\\
$^{2}$ Imperial Centre for Inference and Cosmology (ICIC), Department of Physics, Imperial College, Blackett Laboratory,\\ Prince Consort Road, London SW7 2AZ, U.K.
}

\makeatother

\usepackage{babel}
\begin{document}


\maketitle
\pagerange{\pageref{firstpage}--\pageref{lastpage}} \pubyear{2015}

\label{firstpage} 
\begin{abstract}
When inferring parameters from a Gaussian-distributed data set by computing a likelihood, a covariance matrix is needed that describes the data errors and their correlations. If the covariance matrix is not known a priori, it may be estimated and thereby becomes a random object with some intrinsic uncertainty itself. We show how to infer parameters in the presence of such an estimated covariance matrix, by marginalising over the true covariance matrix, conditioned on its estimated value. This leads to a likelihood function that is no longer Gaussian, but rather an adapted version of a multivariate $t$-distribution, which has the same numerical complexity as the multivariate Gaussian.  As expected, marginalisation over the true covariance matrix improves inference when compared with Hartlap et al.'s method, which uses an unbiased estimate of the inverse covariance matrix but still assumes that the likelihood is Gaussian.
\end{abstract}


\section{Introduction}

A very common problem in statistical inference concerns data that are Gaussian-distributed.  The likelihood of the observed data $\obsboX$ is a multivariate Gaussian, characterised only by a mean data vector $\bomu$ and a covariance matrix $\boSig$:
\begin{equation}
 G(\obsboX | \bomu, \boSig) = \frac{1}{\sqrt{|2\pi \boSig|}}\exp\left[-\half (\obsboX -\bomu)^T\boSig^{-1}(\obsboX - \bomu)   \right].
 \label{Gauss}
\end{equation}
The posterior probability of the parameters is proportional to the likelihood, now treated as a function of the parameters (through the dependence of the mean and the covariance matrix), multiplied by a suitable prior. Ideally one has analytic expressions for the mean and covariance in terms of the model parameters, but in many cases these are not available, and one or both may need to be estimated from simulated data which mimic the experiment that is to be analysed (e.g., \citet{Semboloni,Heymans}), or from the data themselves (e.g., \citet{Budavari}).  However, although an unbiased simulated covariance matrix $\boS$ can be constructed, its inverse is not an unbiased estimator of the inverse (or precision) matrix $\boSig^{-1}$, which is what is needed in the likelihood Eq.~(\ref{Gauss}).  One can construct an unbiased estimator of $\boSig^{-1}$ by a rescaling of $\boS$ \citep{AndersonTW}, as advocated by \cite{Hartlap}. This widens up the credible intervals. If simulations are computationally cheap, then one can generate a large number $N$ of simulated datasets and obtain an accurate estimate of the covariance matrix.  This asymptotic regime occurs only when $N$ far exceeds the size of the data vector, $p$. In many practical cases this is not possible, and the number of simulated datasets is small, with the consequence that statistical noise in the precision matrix propagates into errors in the parameters  \citep{TJK,Dodelson,Hamimeche}. However, there is a more fundamental difficulty with the approach adopted, as it assumes that the likelihood is still Gaussian, albeit with a different precision matrix, whereas in fact it is not. 

A principled way to tackle the problem is to recognise that the simulated data provide {\em samples} of the covariance matrix, so $\boS$ is itself a random object, based on a number of simulations.  For Gaussian data, we have the advantage that the sample distribution of $\boS$ is known, for a given true covariance matrix $\boSig$, and we can exploit this, with a suitable prior, by constructing the probability of $\boSig$ conditional on the sample $\boS$, and then marginalising over the unknown covariance matrix $\boSig$.  This can be done analytically for our preferred choice of Jeffreys prior for $\boSig$.  As a consequence, we properly propagate the uncertainty in the covariance matrix into the final inference, computing the quantity we want, i.e., the likelihood given the {\em simulated} covariance matrix $\boS$ and the number of samples $N$ on which is it based: $P(\obsboX | \bomu, \boS, N)$.  This object, where we keep the dependence on the number of simulated datasets $N$ explicit to emphasize its importance, is the main result of this paper.  It is not Gaussian, but rather follows a modified version of the multivariate $t$-distribution.  In practical terms, it is no more expensive to compute than the Hartlap-scaled Gaussian likelihood, but statistically sound, and can be retrospectively applied to many analyses that have used a different likelihood function by appropriate re-weighting of points, provided that the chains adequately sample the parameter space that the $t$-distribution favours. 

\section{Replacing a true covariance matrix by an estimator}
When inferring cosmological model parameters $\both$ from a data set, we usually have just one p-dimensional observed data vector $\obsboX$, which is a single realization of a statistical process which we assume to be a multivariate Gaussian of which the mean $\bomu$, and the covariance matrix $\boSig$ may depend on the parameters $\both$
\begin{equation}
 \obsboX \sim \mathcal{N}_p\left[\bomu(\both), \boSig(\both)\right].
\end{equation}
In the following, we suppress this dependence on the parameters but it is still implied.

If $\boSig$ were known precisely, the likelihood would be the Gaussian, Eq. (\ref{Gauss}).
However, if $\boSig$ is unknown, and all we have is an estimator $\boS$, then the likelihood $G(\obsboX | \bomu, \boSig)$ must be replaced by another likelihood of which we will show that it is not a Gaussian.

One method - viable for Frequentists - of estimating the covariance matrix, is to draw further independent data vectors from the distribution of $\obsboX$ and to calculate their sample covariance. Typically, such repeated independent measurements are however impossible in cosmology.  Nonetheless, if we can simulate the observation, then we are able to generate further samples $\boXim_i;\quad i=1,\ldots N$, that are statistically equivalent to the single observation $\obsboX$. The covariance matrix $\boS$ can then be estimated from these simulations, and the likelihood that we require is the probability of the data, given $\boS$ and the number of simulations on which it is based, i.e., $P(\obsboX |\bomu, \boS, N)$. 

If we run $N$ independent simulations, then $ \boXimbar = \frac{1}{N}\sum_{i = 1}^N \boXim_i$
is the average, and an unbiased estimator of $\boSig$ is
\begin{equation}
 \boS = \frac{1}{N-1} \sum_{i = 1}^N  (\boXim_i - \boXimbar)(\boXim_i - \boXimbar)^T.
\end{equation}
In the following subsection, we derive an analytical replacement for the Gaussian likelihood, Eq.~(\ref{Gauss}), conditioned on an estimate $\boS$, and from Sect. \ref{Sect:Gauss_vs_T} onwards we study the effects of this replacement on parameter inference.  

\subsection{Derivation of the multivariate t-distribution}
We now derive the likelihood $P(\obsboX |\bomu, \boS, N)$ that depends on an estimator $\boS$ instead of the true covariance $\boSig$. 

Any matrix of the type $\sf{M} = \sum_{i = 1}^m \boldsymbol{Y}_i \boldsymbol{Y}_i^T$ is by construction a Wishart matrix \citep{Wishart,MardiaKentBibby,AndersonTW}, if $\boldsymbol{Y}$ is drawn from a multivariate Gaussian. When estimating a covariance matrix by averaging over random samples drawn from simulations, the estimated covariance matrix is a Wishart matrix, and has a Wishart distribution \citep{AndersonTW},
\begin{equation}
 \calW(\boS | \boSig/n, n ) = \frac{|\boS|^\frac{n-p-1}{2} \exp\left[ -\half n\trace{\boSig^{-1} \boS }\right]    }{2^\frac{pn}{2} |\boSig/n|^\frac{n}{2} \Gamma_p\left( \frac{n}{2} \right)}
 \label{Wishart}
\end{equation}
where we call $n = N-1$ the degrees of freedom and $\Gamma_p$ is the $p$-dimensional Gamma function.
By the central limit theorem, this distribution is also asymptotically appropriate if the sampling distribution of $\boX$ is non-Gaussian.

We can invert this distribution to yield the distribution $P(\boSig | \boS, N)$ of the true covariance matrix $\boSig$ conditioned on the estimator $\boS$, by using Bayes' Theorem 
\begin{equation}
P(\boSig | \boS, N) \pi(\boS) =  \calW(\boS | \boSig/n,n) \pi(\boSig) 
\end{equation}
and adopting priors $\pi$. Since the determinant of the positive-definite covariance matrix is strictly positive, it is a scaling parameter, and we therefore assume the independence-Jeffreys prior \citep{Jeffreys,SunBerger}
\begin{equation}
 \pi(\boSig) \propto |\boSig|^{-\frac{p+1}{2}}.
\end{equation}
This is by construction invariant under reparametrizations, and can therefore be regarded as uninformative, independent of the choice of parameters.\footnote{The power $(p+1)/2$ also leads to $N-1$ degrees of freedom in the inverse Wishart distribution, which is an intuitive result. Another power would only change the degrees of freedom, showing that the influence of the prior can be lessened by increasing the number of simulations $N$.} We then have
\begin{eqnarray}
P(\boSig | \boS, N) & \propto &\calW(\boS | \boSig/n, n  )\pi(\boSig ) \nonumber\\
&\propto& |\boSig|^{-\frac{n+p+1}{2}} \exp\left[-\half n \trace{\boSig^{-1} \boS} \right]\nonumber\\
&\propto &\calW^{-1}(\boSig | n\boS,n  )
\end{eqnarray}
showing that the uncertainty of the unknown true $\boSig$ can be described by an inverse Wishart distribution, conditioned on the sample estimate,
\begin{equation}
 \calW^{-1}(\boSig | \boC, n) =  \frac{|\boC|^{\frac{n}{2}} |\boSig|^{-\frac{n+p+1}{2}}  \exp\left(-\half  \trace{\boSig^{-1} \boC} \right)}{2^\frac{np}{2}  \Gamma_p\left(\frac{n}{2}\right)   }
 \label{InverseWishart}
\end{equation}
where we used $\boC = n \boS$. Increasing the estimates, $N=n+1$, of the covariance matrix, will make this distribution more sharply peaked, reflecting the improvement of the estimation. 

Given the distribution Eq.~(\ref{InverseWishart}), we can now marginalize the Gaussian likelihood over the unknown covariance, to find what we are after, which is the likelihood of the data $\obsboX$, given a mean $\bomu$ and an estimate $\boS$ of the covariance matrix from $N$ simulations:
\begin{eqnarray}
 & &P(\obsboX |\bomu, \boS, N)  = \int\,\mathd \boSig\,  G(\obsboX | \bomu, \boSig) P(\boSig | \boS, N) \nonumber\\
 &\propto &\int\,\mathd \boSig\, |\boSig|^{-\frac{N+p+1}{2}}\exp\left[ -\half \trace{\boSig^{-1} {\sf{Q}}}\right]
\end{eqnarray}
where we have defined ${\sf{Q}} = n\boS +(\obsboX -\bomu)(\obsboX - \bomu)^T$. The last line is structurally the integration over an unnormalized inverted Wishart distribution $\calW^{-1}(\boSig | {\sf{Q}},N)$, so the result is the normalization constant as in Eq.~(\ref{InverseWishart}), leading to
\begin{equation}
P(\obsboX |\bomu, \boS, N) \propto |{\sf{Q}}|^{-\frac{N}{2}}.
\end{equation}
Resubstituting ${\sf{Q}}$, using the matrix identity
\begin{equation}
 |\sf{A} + \boldsymbol{b}\boldsymbol{b}^T | = |\sf{A}|(1+\boldsymbol{b}^T\sf{A}^{-1}\boldsymbol{b})
\end{equation}
and normalizing, we arrive at the likelihood for the $p$-dimensional dataset $\obsboX$, conditioned on the mean $\bomu$ and a sample of the covariance matrix $\boS$ from $N$ simulations: 
\begin{equation}
P(\obsboX | \bomu, \boS, N) = 
\frac{\bar{c}_p |\boS|^{-1/2}}{\left[ 1 + \frac{(\obsboX -\bomu)^T\boS^{-1} (\obsboX - \bomu) }{N-1}\right] ^{\frac{N}{2}}}
  \label{cosmo_tdistrib}
\end{equation}
This is a cosmologist's version of a multivariate $t$-distribution. It is not the standard (Frequentist) multivariate $t$-distribution, which jointly estimates the mean and its covariance from a data set of $N$ data vectors. In contrast, we have assumed exactly one data vector that determines where the likelihood will peak - and $N$ simulated vectors from which we estimate the covariance. 
The normalization is
\begin{equation}
 \bar{c}_p = \frac{\Gamma\left(  \frac{N}{2}   \right)}{ \left[\pi (N-1)\right]^{p/2} \Gamma \left( \frac{N-p}{2} \right) },
 \label{cbar}
\end{equation}
where $\Gamma$ is the usual Gamma function and we require $N > p$. For expensive simulations, when a feasible $N$ is still comparable to $p$, the differences between a Gaussian and the $t$-distribution become important. So if a covariance matrix must be replaced by an estimator from simulations, the modified $t$-distribution Eq.~(\ref{cosmo_tdistrib}) replaces the multivariate Gaussian Eq.~(\ref{Gauss}). This is the main result of the paper.

\section{Attempting to debias a Gaussian likelihood}
\label{Sect:Hartlap}

\begin{figure}
\includegraphics[width=0.47\textwidth]{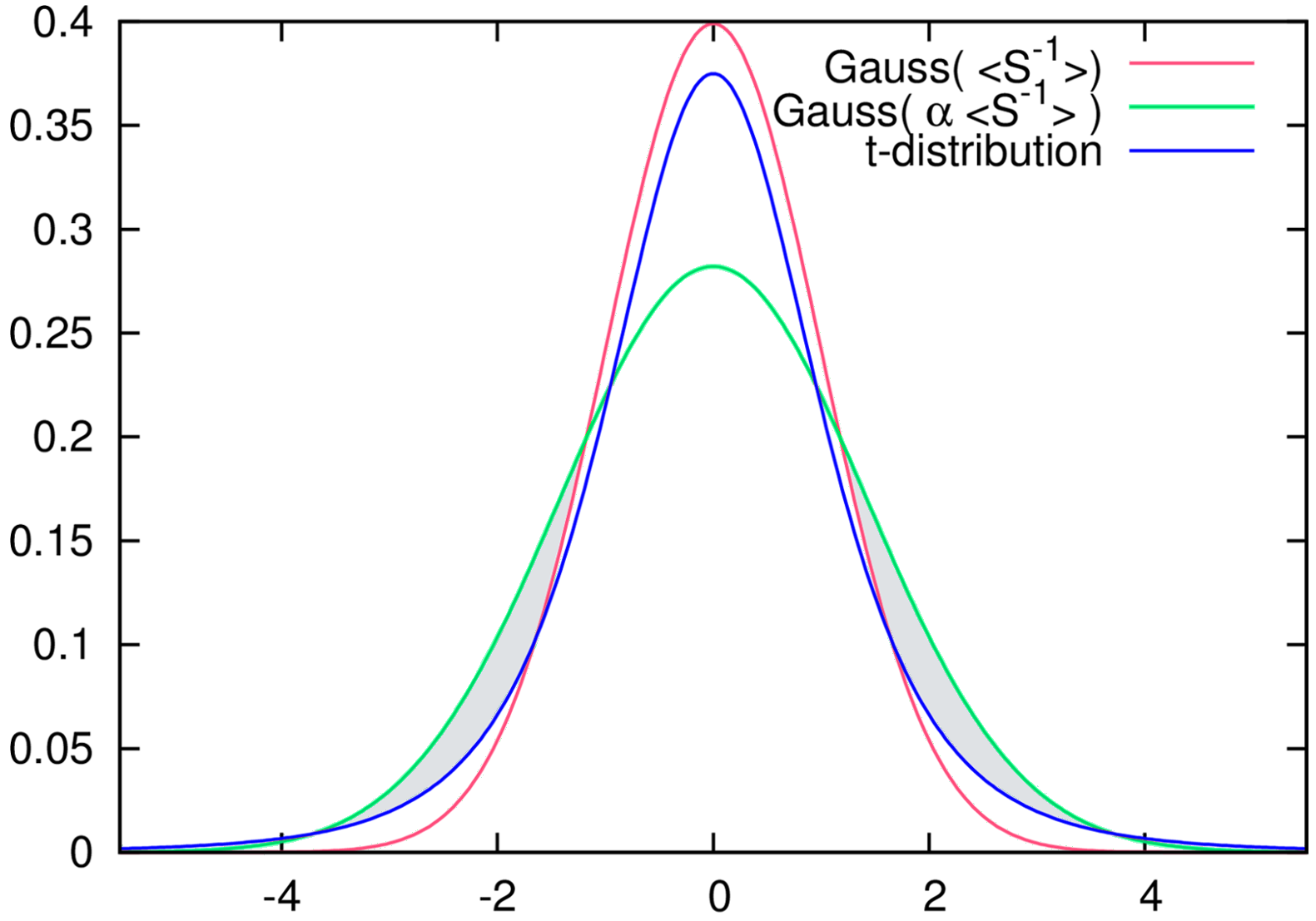} 
\caption{Comparison of the two Gaussian likelihoods and the $t$-distribution for a {\em particular} estimated $\boS$, using $N = 5, p = 1, \alpha = 0.5$ which are  examples. The grey shaded areas indicate the heavy and short wings of the Hartlap-scaled likelihood.}
\label{Likelihoods}
\end{figure}

\begin{figure}
\includegraphics[width=0.47\textwidth]{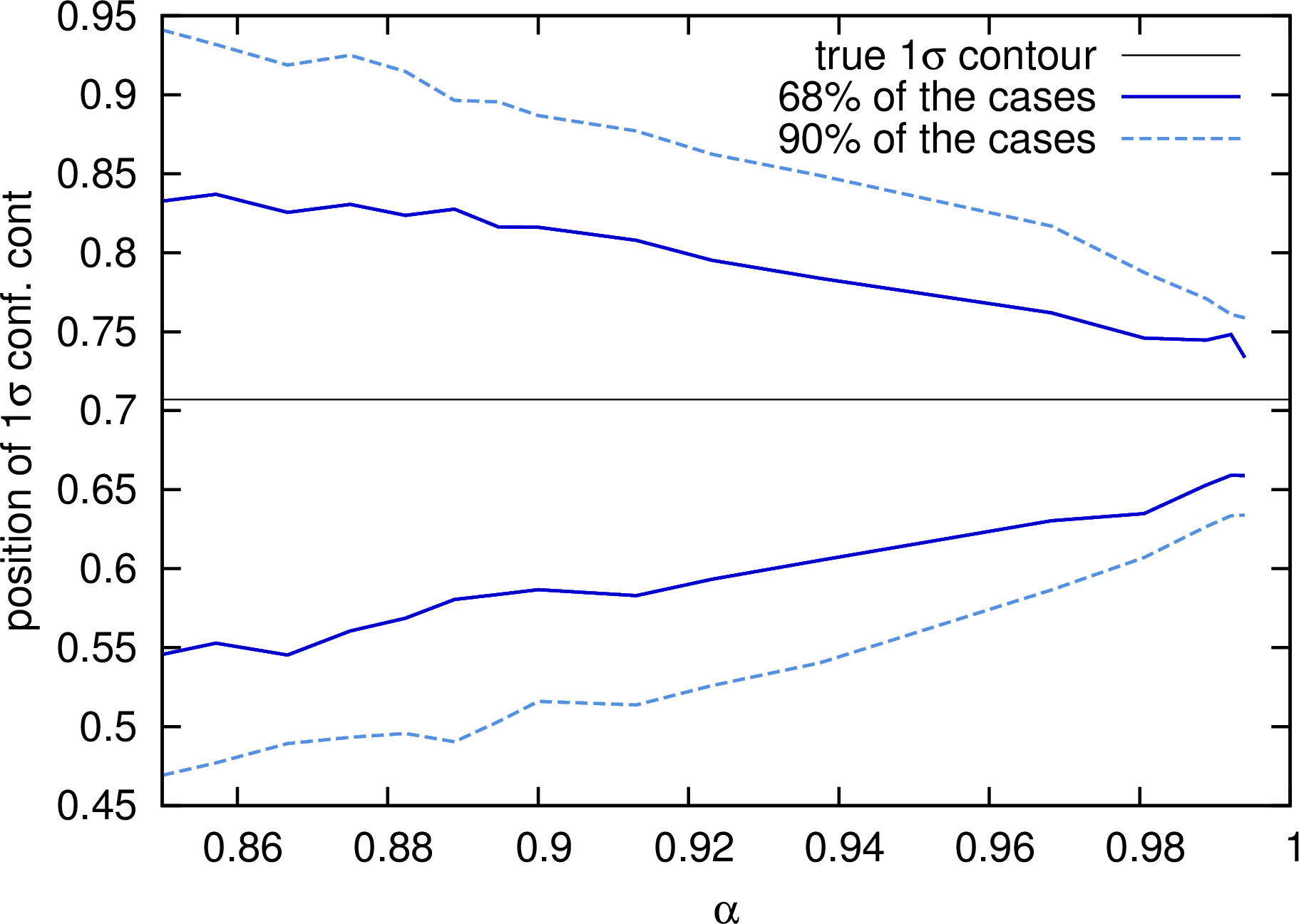} 
\caption{The $1\sigma$-confidence contour of a onedimensional normal distribution lies at $1/\sqrt{2}\approx 0.707$. However, if the covariance is estimated from simulations, its random scatter will make the estimated likelihood randomly too narrow or too broad. In $68\%$ $(90\%)$ of the estimated covariances, the then deduced $1\sigma$-contour falls into the area bordered by the dark blue (dashed blue) lines. The number of simulations increases with $\alpha$ from Eq.~(\ref{alpha}).
}
\label{Conficonts}
\end{figure}

Instead of using the $t$-distribution Eq.~(\ref{cosmo_tdistrib}) it has become standard in cosmology to follow a procedure outlined by \citet{Hartlap}, where the authors propose to stick with a Gaussian likelihood, and only to replace the true inverse covariance matrix by a scaled inverse sample covariance matrix
\begin{equation}
 \boSig^{-1} \rightarrow \alpha \boS^{-1}
\label{proposal}
\end{equation}
with 
\begin{equation}
 \alpha = \frac{N-p-2}{N-1}.
\label{alpha}
\end{equation}
This is motivated by the fact that $\boS^{-1}$ follows an inverse Wishart distribution, which has a biased expectation value 
$\langle \boS^{-1} \rangle = \alpha^{-1}\boSig^{-1}$
as shown in \citet{AndersonTW}. Here, the angular brackets denote averaging over the inverse Wishart distribution.

\citet{Hartlap} argue that this debiased inverse covariance matrix will remove all biases from parameter inference. However, the situation is more complex. In a Bayesian anlaysis one would not necessarily define an estimator $\hat{\both}$, but if one does, the bias is
$b_\theta = \langle \hat{\both} \rangle -\both$ where the angular brackets now denote the average over the likelihood of the parameters.   Adopting the wrong sampling distribution will yield incorrect posterior distributions, with biased parameter estimates (should they be made) and incorrect errors, even if the inverse covariance matrix itself has been debiased.

We compare univariate examples of the likelihoods and the modified $t$-distribution Eq.~(\ref{cosmo_tdistrib}) in Fig.~\ref{Likelihoods}: the Hartlap-scaled and the unscaled Gaussian only differ in width, whereas the $t$-distribution has a more sharply peaked central region but broader extreme wings than a Gaussian, allowing for more scatter away from the peak. 

Additionally, the scaling in Eq.~(\ref{proposal}) implies a sharp mapping between the estimator $\boS^{-1}$ and $\boSig^{-1}$, which does not account for the randomness of $\boS^{-1}$, due to the finite width of the inverse Wishart distribution. Therefore, $\alpha \boS^{-1}$ applied to a \emph{single} given $\boS^{-1}$ should not be interpreted as a reliable `debiasing' but rather a scaling that widens up the Gaussian likelihood Eq.~(\ref{Gauss}) in an essentially random way. This randomness will propagate through the parameter inference and introduce a scatter of the likelihood contours of which we show a simple example in Fig.~\ref{Conficonts}. This scatter can only be reduced by estimating the inverse covariance matrix more precisely, see also \citep{TJK,Dodelson}.

\section{Comparison of the distributions}
\subsection{Illustrative univariate example}
\label{Sect:Gauss_vs_T}

\begin{figure}
\includegraphics[width=0.47\textwidth]{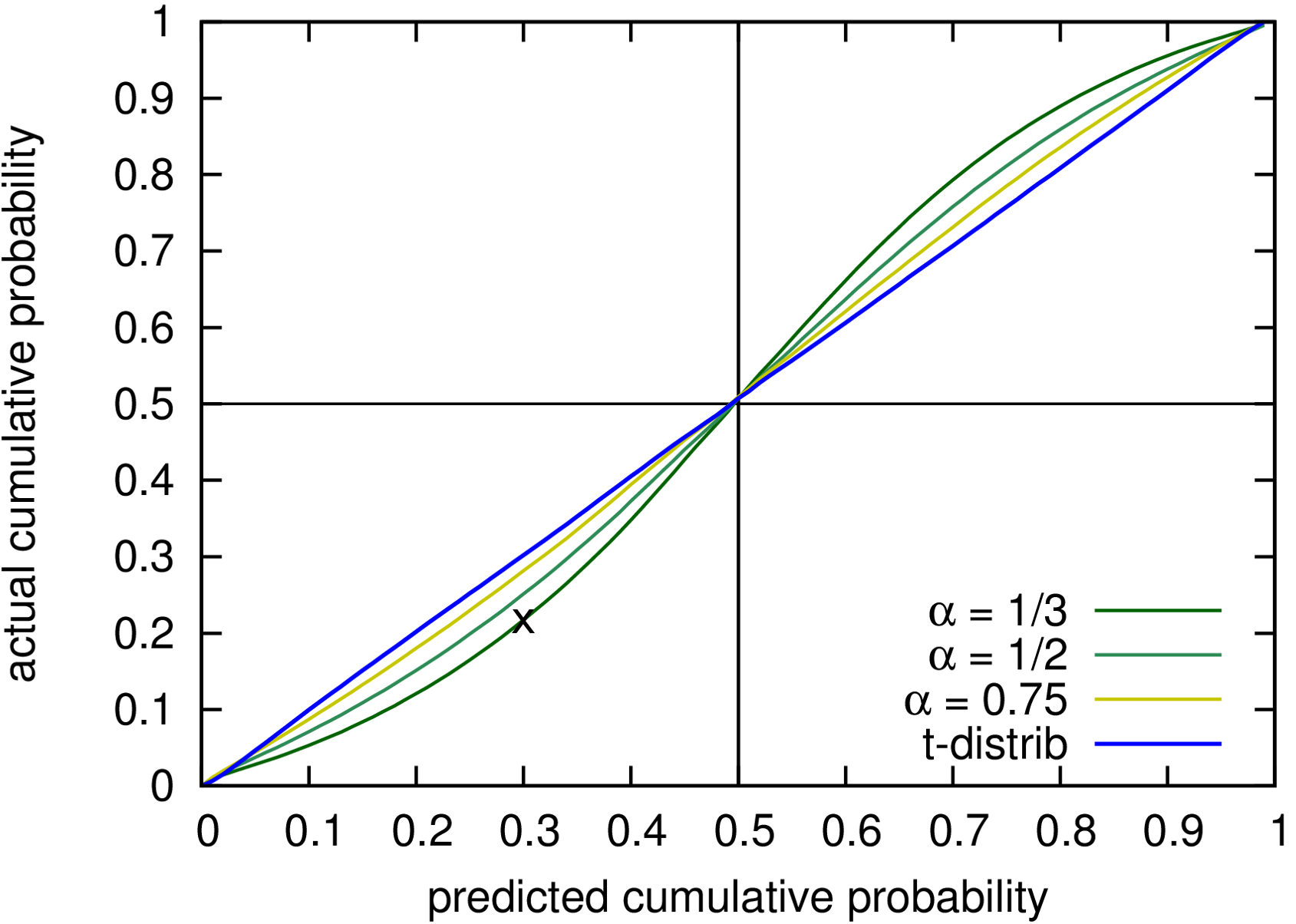} 
\caption{Predicted versus true cumulative probability for an illustrative univariate estimation of a mean. The $t$-distribution follows the diagonal line of unit slope, meaning it predicts correctly the shape of the likelihood, whereas the Hartlap-scaled Gaussian is too broad. For example, the marked point is the lower 30\%-tail of the Hartlap-scaled Gaussian -- but in reality the true mean falls into this tail only with a probability of 0.2.}
\label{Comparison}
\end{figure}

We illustrate with a univariate frequentist example that the Hartlap-scaled Gaussian introduces errors into the parameter inference, whereas Eq.~(\ref{cosmo_tdistrib}) does not. We choose a \emph{true} mean $\mu_t$, which we want to estimate in the following. We then produce 10,000 Gaussian data sets with this mean, and produce 150 estimates of the covariance matrix from $N$ further samples (where $N$ determines $\alpha$). For each data set and each covariance matrix, we then calculate the Hartlap-scaled likelihood and the modified $t$-distribution. Both the Hartlap-scaled Gaussian and the $t$-distribution of $\mu_t$ make quantitative predictions such as stating that $\mu_t$ will fall 5\% of the time into the lower 5\% tail of the likelihood, or 68\% of the time into the 68\% likelihood contour, given some data sets. But since the two likelihoods differ in shape, their lower-tail probabilities and likelihood contours will also differ, and only one will make the correct quantitative predictions. Since we know the true mean, we can test this. Likelihood contours and tail-probabilities can be converted into each other, so it is sufficient to test only one. We choose the lower $x$-percent tail probability, i.e. the cumulative probability function and check whether the $x$-percent cumulative distribution does indeed cover the true mean $x$ percent of the times. In Fig.~\ref{Comparison}, we find that only the $t$-distribution correctly reproduces the cumulative distribution - the line is straight with a slope of unity. The Hartlap-scaled Gaussian does not capture the scatter around the peak correctly, which will lead to a misestimate of the parameter errors, even on average. As expected, the discrepancy decreases as more simulations are included in the estimation of $\boS$ (i.e., as $\alpha\rightarrow 1$).

\subsection{Assessment of confidence in higher dimensions}

\begin{figure*}
\includegraphics[width=\textwidth]{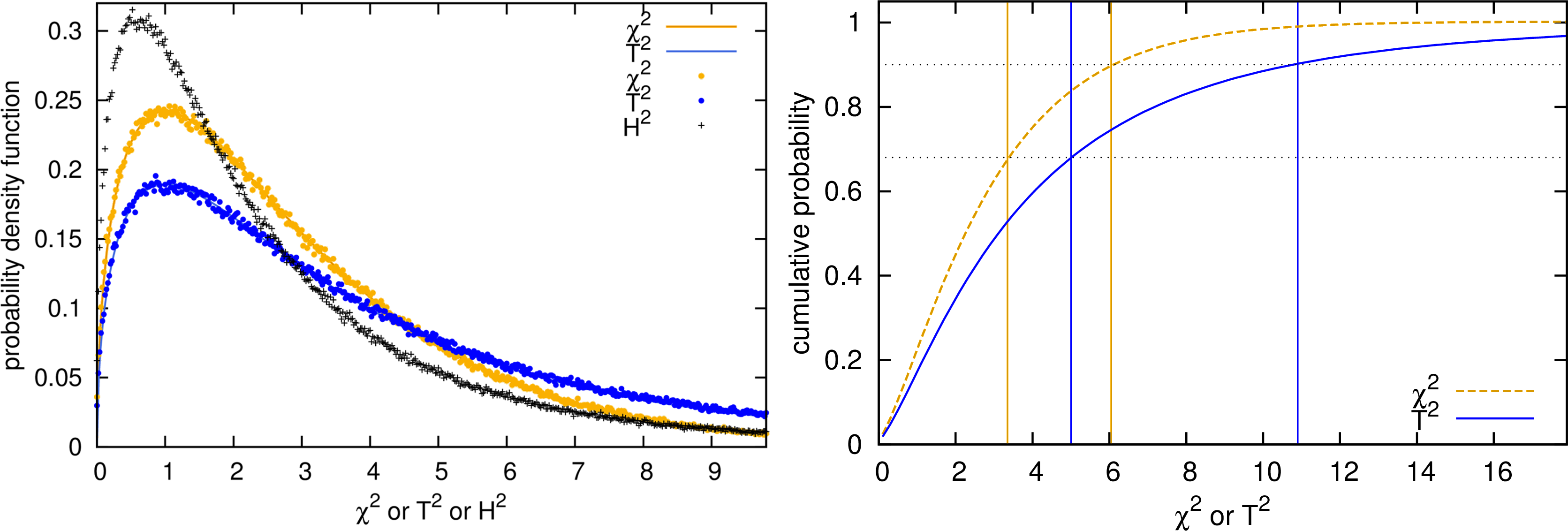} 
\caption{Left: The distribution of different interpretations for $(\obsboX-\bomu)^T \boS^{-1}(\obsboX-\bomu)$, using $p = 3$, $N = 10$. Dots represent simulations, solid lines are the analytical $\chi_p^2$- and $T^2$-distribution. For $N \gg p$, the $T^2$-distribution approximates the $\chi_p^2$-distribution. The closer $N$ is to $p$, the more differs the $T^2$-distribution from the $\chi_p^2$-distribution, being broader than $\chi_p^2$, leading to a cumulative distribution that rises more slowly. The Hartlap-scaled $H^2$ follows the black distribution which is more sharply peaked than the $\chi_p^2$, although the $\chi_p^2$-distribution is the minimal scatter that one can achieve; this means the Hartlap-scaled $H^2$ underestimates the joint scatter of $\obsboX$ and $\boS^{-1}$.
Right: The cumulative distributions of $\chi^2$ and $T^2$ from the left. The vertical lines mark the $68\%$ and $90\%$ confidence limits.}
\label{DiffDists}
\end{figure*} 

The issue at hand can be studied in higher dimensions by investigating the distribution of the following quantities:
\begin{equation}
\chi^2 = \left(\obsboX -  \bomu \right)^T\boSig^{-1}\left(\obsboX -\bomu \right)
\end{equation}
which is the true $\chi^2$; the same quantity but with the estimated $\boS$ replacing $\boSig$,
\begin{equation}
 T^2 = \left(\obsboX -  \bomu \right)^T\boS^{-1}\left(\obsboX -\bomu \right);
\label{T2}
\end{equation}
and the Hartlap-scaled version
\begin{equation}
 H^2 = \left(\obsboX -  \bomu \right)^T\alpha \boS^{-1}\left(\obsboX -\bomu \right).
\end{equation} 
By construction, we have $\langle H^2\rangle = \langle \chi^2 \rangle$, meaning the Hartlap-scaling does indeed debias the expectation value. It does however underestimate statistical scatter, as we shall show in the following.

$\chi^2$ follows the $\chi_p^2$-distribution, which only arises if the covariance is precisely known and indeed the correct covariance of $\obsboX$.
The quantity $T^2$ will not follow the $\chi^2_p$-distribution, because it contains not only a random vector $\obsboX \sim \mathcal{N}_p(\bomu, \boSig)$, but additionally the random estimate of the covariance matrix that follows the Wishart distribution $\calW(\boSig/n,n)$. $T^2$ therefore follows
\begin{equation}
 \frac{T^2  (n-p+1)}{pn} \sim F_{p,n-p+1}
\end{equation}
where $n = N-1$, and  the $F_{p,n-p+1}$ is the F-distribution of $p$ and $n-p+1$ degrees of freedom \citep{AndersonTW}.  Consequently, a change of variables shows that,
\begin{equation}
 T^2 \sim \frac{\Gamma\left(\frac{n+1}{2} \right)}{\Gamma(p/2)\Gamma[(n-p+1)/2]} \frac{n^{-p/2}(T^2)^{p/2-1} }{(T^2/n +1)^{\frac{n+1}{2}}}.
\label{T2distrib}
\end{equation}
instead of $T^2 \sim \chi_p^2$, see Fig.~\ref{DiffDists}. Only for $N \rightarrow \infty$ will the Wishart distribution tend towards a delta-function, and the distribution of $T^2$ will then tend towards a $\chi^2_p$-distribution.

The distribution of the Hartlap-scaled $H^2$ is more sharply peaked than that of $\chi^2$, thereby suggesting that the experiment has less statistical scatter than the $\chi_p^2$ distribution on average. This is  impossible since the $\chi_p^2$ distribution is subject to scatter of the random vector $\obsboX$ only.

The cumulative probabilities $P_{\rm c}(\chi_{\rm c}^2)$ or $P_{\rm c}(T_{\rm c}^2)$ give our confidence that the mean $\bomu$ of the multivariate vector $\obsboX$ is enclosed within an ellipsoid bounded by $\chi_{\rm c}^2$ or $T_{\rm c}^2$. The more slowly rising cumulative distribution function of $T^2$ therefore shows that we need $T^2 > \chi^2$ in order to achieve the same confidence that the mean is captured within the confidence contours. In parameter space, this will lead to an increase of the Bayesian confidence intervals.

\subsection{Reweighting an MCMC chain that sampled from a Gaussian likelihood.}
We have shown above that $T^2$, $\chi^2$ and $H^2$ follow different distributions, which will affect parameter inference. Often, the error of confusing a $T^2$ with a $\chi^2$ or $H^2$ can retrospectively be undone with very little numerical effort by reweighting an existing MCMC chain.

\begin{figure}
\includegraphics[width=0.47\textwidth]{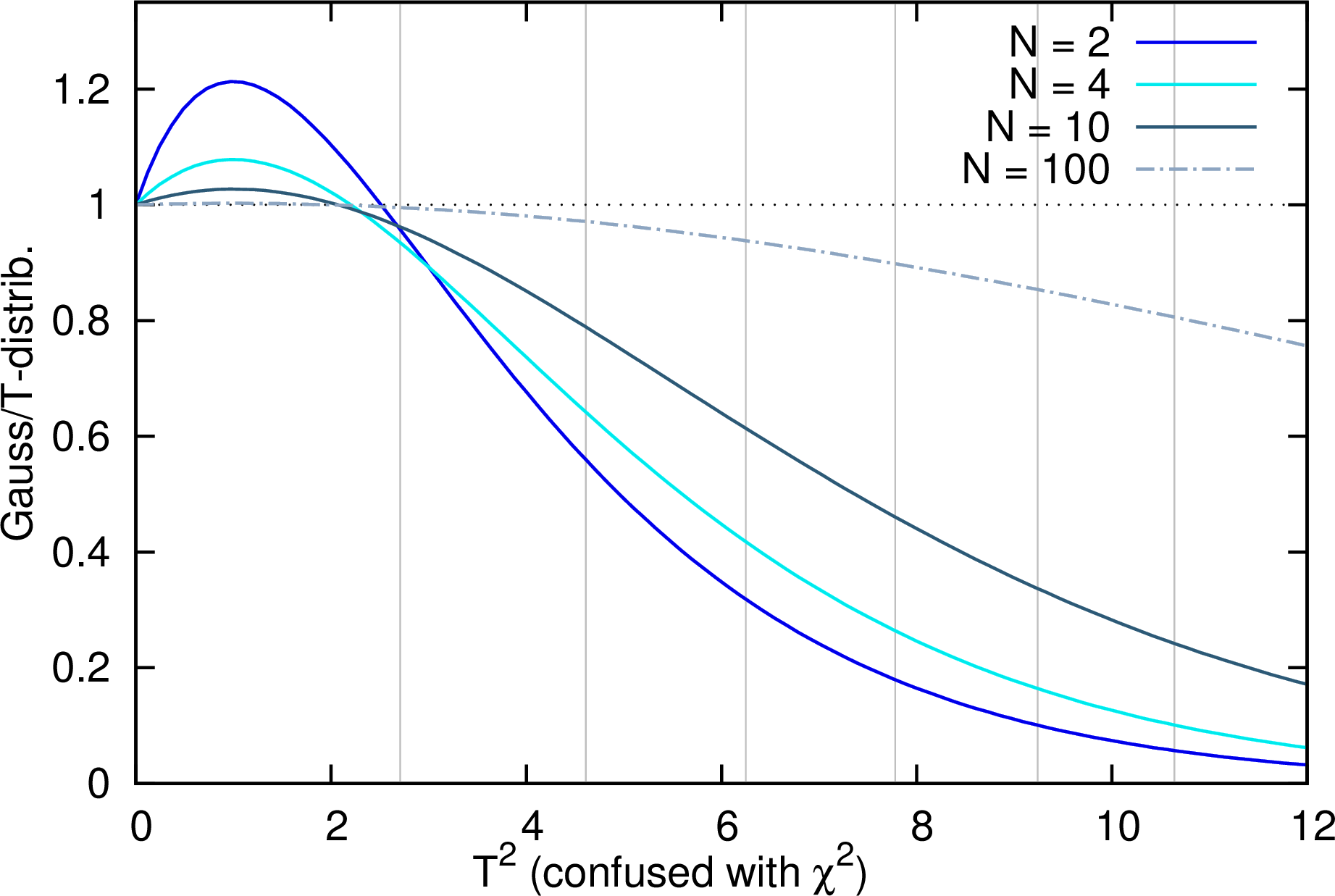} 
\caption{Unnormalized weights $G(\obsboX, \bomu, \boS^{-1})/P(\obsboX, \bomu, \boS^{-1},n)$ for mapping between a Gaussian likelihood and a $t$-distribution. The normalization depends on the dimensionality of the data set, and leads to an offset along the $y$-axis, that is however independent of theoretical parameters. The number of simulations in the covariance matrix is $N$. The vertical lines depict the $\chi^2$ values (2.71,4.61,6.25,7.78,9.24,10.64) that enclose $90\%$ confidence for a multivariate Gaussian.}
\label{Fig:ChiSquare}
\end{figure}

In Fig \ref{Fig:ChiSquare} we plot weights for reweighting a chain that sampled from $\exp(-\chi^2/2)$. If a Hartlap-scaling has been applied, it would additionally need to be removed. 

We note that the maximum of the $t$-distribution in the full parameter space coincides with the maximum of $\chi^2$ (and also of $H^2$), but once any parameters are marginalised over, the resulting parameter posteriors will not in general peak in the same place.

\section{Conclusions}
We have studied how statistical uncertainties in an estimated covariance matrix affect parameter inference. We summarize our findings as follows.

For data $\obsboX$ drawn from a multivariate Gaussian, the likelihood will be Gaussian if the data covariance $\boSig$ is exactly known. If however the covariance is estimated from $N$ simulations, $\boS = 1/(N-1)\sum_{i = 0}^N (\boX_i -\bar{\boX})(\boX_i -\bar{\boX})^T$, the estimator $\boS$ is unbiased, but $\boS^{-1}$ is not an unbiased estimator of $\boSig^{-1}$. An unbiased estimator is $\alpha \langle \boS^{-1}\rangle = \boSig^{-1}$ where $\alpha = (N-p-2)/(N-1)$ \citep{AndersonTW}. An earlier proposal, by \cite{Hartlap}, uses the unbiased estimate $\alpha \boS^{-1}$ of the inverse covariance matrix, but keeping a Gaussian likelihood. The statistical scatter of the estimator 
$\boS^{-1}$ is not fully accounted for, and this yields posteriors that are on average simultaneously too broad in their centres, yet not broad enough in the extremes.

The principled approach is to recognise that we have a {\em sample} of the covariance matrix $\boS$, and compute the likelihood by marginalising over the  inverse-Wishart distribution of the true covariance matrix $\boSig$, conditioned on  $\boS$.  This gives a modified multivariate $t$-distribution $P(\boX_{\rm o}|\bomu,\boS,N)$, given by Eq. (\ref{cosmo_tdistrib}). This is what we require for parameter inference and is the main result of this paper. 

For parameter inference in the presence of a covariance matrix estimated from a finite number of simulations, our results imply that MCMC chains should evaluate the modified $t$-distribution Eq.~(\ref{cosmo_tdistrib}) at each sample point, instead of a Gaussian distribution. The numerical complexity will not be increased by this. It stays constant since both distributions must evaluate the quantity $(\obsboX -\bomu)^T\boS^{-1}(\obsboX - \bomu)$. Consequently, a reweighting of existing MCMC chains is possible without much effort if the chains record $(\obsboX -\bomu)^T\boS^{-1}(\obsboX - \bomu)$.

\section{Acknowledgements}
ES acknowledges financial support from the RTG \emph{Particle Physics beyond the Standard Model}, through the DFG fund 1940 and the transregional collaborative research centre TR 33 `\emph{The Dark Universe}' of the DFG.  We thank Andrew Jaffe, Justin Alsing and Ewan Cameron for useful discussions and comments.

\bibliographystyle{mn2e}
\bibliography{TDist}

\label{lastpage} 
\bsp 
\end{document}